\documentclass[doublecol]{epl2} 
\usepackage{amsmath}
\usepackage{mathtools}
\usepackage{amssymb}
\usepackage{amsthm}
\usepackage{amsfonts}
\usepackage{hyperref}

\usepackage{graphicx}
\usepackage{color}
\usepackage[percent]{overpic}
\usepackage{enumerate}
\usepackage[export]{adjustbox}
\allowdisplaybreaks

\newcommand{\ket}[1]{|\,{#1}\,\rangle}

\newcommand{\expec}[1]{\langle #1 \rangle}

\newcommand{\sub}[2]{{#1}_{\mbox{\!\! \scriptsize #2}}}

\def\beq{\begin{equation}}
\def\eeq{\end{equation}}

\def\CR{\nonumber\\[0.15cm]}

\newcommand{\rref}[1]{Ref.~\cite{#1}}
\newcommand{\fref}[1]{Fig.~\ref{#1}}
\newcommand{\frefp}[2]{Fig.~\ref{#1}~(#2)}

\newcommand{\eref}[1]{Eq.~(\ref{#1})}
\newcommand{\esref}[2]{Eqs.~(\ref{#1}) and (\ref{#2})}

\newcommand{\cref}[1]{chapter~\ref{#1}}

\newcommand{\Cref}[1]{Chapter~\ref{#1}}

\newcommand{\bref}[1]{(\ref{#1})}

\usepackage{ulem}  
\title{Solitary waves explore the quantum-to-classical transition}
\shorttitle{Solitary waves explore the quantum-to-classical transition}

\author{A. Sreedharan \inst{1} \and S Kuriyattil \inst{1} \and S. Choudhury\inst{1,2} \and R. Mukherjee \inst{1,3} \and A. Streltsov \inst{4,5} \and S.~W\"uster \inst{1}  }
\shortauthor{A. Sreedharan \etal} 

\institute{                    
  \inst{1} Department of Physics, Indian Institute of Science Education and Research (IISER), Bhopal, Madhya Pradesh 462066, India\\
   \inst{2} New Zealand Institute for Advanced Study and Centre for Theoretical Chemistry and Physics, Massey University, Auckland 0632, New Zealand\\
  \inst{3} Department of Physics, Imperial College, SW7 2AZ, London, UK \\
  \inst{4} Theoretische Chemie, Physikalisch-Chemisches Institut, Universit{\"a}t Heidelberg, Im Neuenheimer Feld 229, D-69120 Heidelberg, Germany \\
  \inst{5} SAP Deep Learning Center of Excellence and Machine Learning Research 
SAP SE, Dietmar-Hopp-Allee 16, 69190 Walldorf, Germany
}

\abstract{ It is an open fundamental question how the classical appearance of our environment arises from the underlying quantum many-body theory. We propose that phenomena involved in the quantum-to-classical transition can be probed in collisions of bright solitary waves in Bose-Einstein condensates, where thousands of atoms form a large compound object at ultra cold temperatures. 
For the experimentally most relevant quasi-1D regime, where integrability is broken through effective three-body interactions, we find that ensembles of solitary waves exhibit complex interplay between phase coherence and entanglement generation in beyond mean-field simulations using the truncated Wigner method:
An initial state of two solitons with a well defined relative phase looses that phase coherence in the ensemble, with its single particle two-mode density matrix exhibiting similar dynamics as a decohering two mode superposition. This \emph{apparent} decoherence is a prerequisite for the formation of entangled superpositions of different atom numbers in a subsequent soliton collision. The necessity for the solitons to first decohere is explained based on the underlying phase-space of the quintic mean field equation. We show elsewhere that superpositions of different atom numbers later further evolve into spatially entangled solitons. Loss of ensemble phase coherence followed by system internal entanglement generation appear in an unusual order in this closed system, compared to a typical open quantum system.
}
\begin{document}
\maketitle

\section{Introduction}
Why most of the world around us follows the classical laws of physics, while being built from quantum mechanical microscopic constituents, is a paramount puzzle of modern physics \cite{Schlosshauer_decoherence_review}. As experiments are pushing towards superposition states with more and more constituents \cite{friedman:squidcat,Gao:hyperentangle,Leibfried:hyperfinecat,takahashi:ancilla,gerlich:organicmols,lu:graphstates,monroe:singleatom,Arndt_C60_doubleslit,arndt_hornberger_limits,Eibenberger_molinterf_pccp}, all points to a central role of decoherence and system-environment entanglement in the transition from quantum to classical appearance \cite{Schlosshauer_decoherence_review,hackermueller_thermaldecoh,Hornberger_collisionaldecoh}. These both inherently rely on the ease with which entanglement proliferates in quantum many body systems.

In ultra-cold gaseous Bose-Einstein condensates (BEC), thousands of atoms can form a compound object, a bright soliton \cite{footnote_solitons}, due to their weakly attractive contact interactions. These solitons are localized solutions of the Gross-Pitaevskii equation (GPE) that governs the mean field of the condensate \cite{book:solitons,li_rev}. They are protected from dispersion by the non-linearity of atomic interactions, and regularly created since 2002 \cite{khay:brighsol,strecker:brighsol,li_rev,gap_exp,jila:solitons,Nguyen_modulinst,Nguyen_solcoll_controlled,Marchant_Quantrefl,Marchant_controlledform,Medley_evapsoliton,Lepoutre_sol_Ka,McDonald_solitoninterf,Everitt_modinst,Mesnarsic_cesiumsol_PhysRevA,sanz2022interaction,carli_bmw}, motivated by fundamental studies and atom interferometry \cite{Mesnarsic_cesiumsol_PhysRevA,Tsarev_metrology_optexpress,Tsarev_metrology_NJP2019,helm_bmw,wales_bmw,Martin_bsw,Helm_barrier_QCT_PhysRevA.89.033610}.

Quantum soliton \cite{Drummond_squeeze,drummond_noise} collisions in strictly one dimension (1D) can be studied using the integrable Lieb-Liniger-MacGuire (LL) model \cite{LL_model_PR}, 
which prohibits atom transfer between the colliding solitons, since an initial set of single particle quasi-momenta cannot be changed by collisions \cite{lai_quantsol_I,lai_quantsol_II}. However,   experiments to date are in the quasi-1D (Q1D) regime, where transverse dynamics remains important for atomic scattering. The integrability of the LL model is then broken through effective three-body interactions, due to virtual transitions of atoms to transverse modes \cite{Muryshev_darksolelong_quintic_PRL,Sinha_solfriction_quintic_PRL,Mazets_breakinteg_PhysRevLett}.

\begin{figure}[htb]
\centering
\includegraphics[width=0.99\columnwidth]{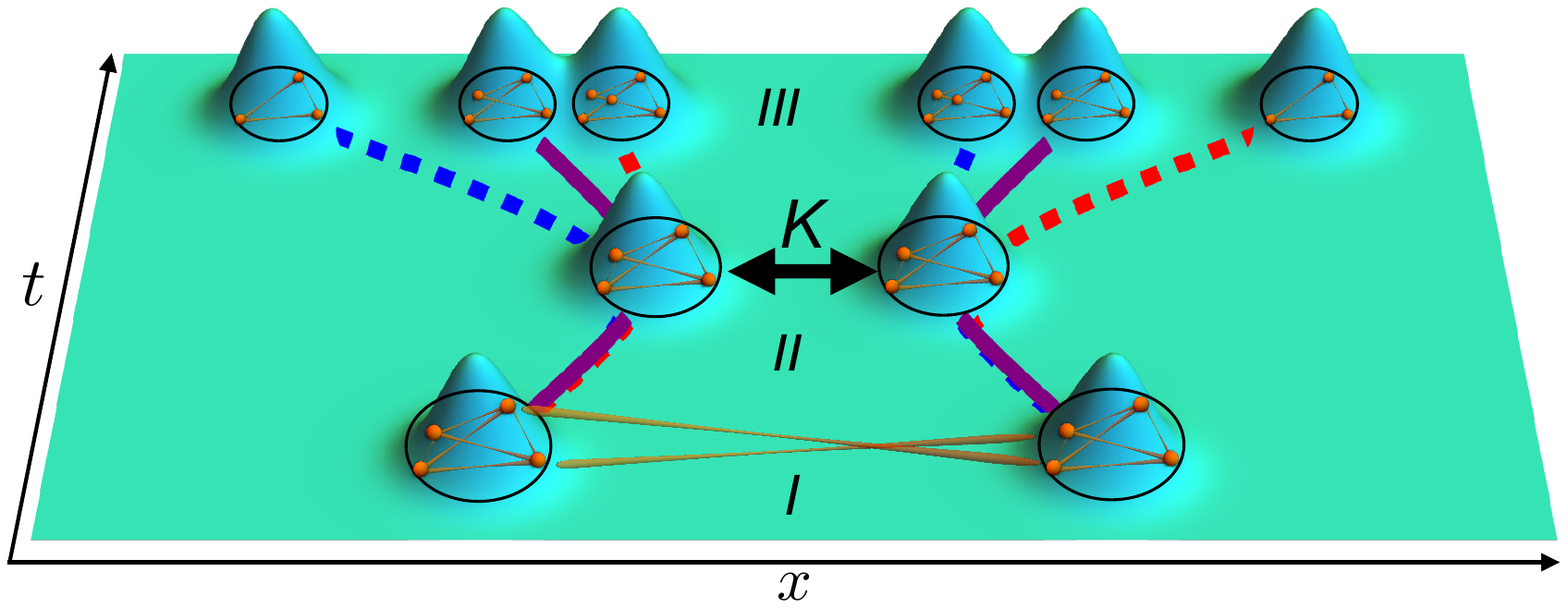}
\caption{\label{sketch} Interplay of coherence and entanglement generation during soliton collisions. Initially, two separated solitons loose coherence (brown stripes) in the ensemble due to phase diffusion (from I to II). \emph{Only after} this fragmentation will a  collision cause significant atom transfer $K$ due to integrability breaking in three dimensions. This subsequently 
gives rise to a quantum state in which atom number, momentum and position of both solitons are entangled (matching colored lines towards III).
} 
\end{figure}
 Here, we show that some properties of solitons and soliton collisions can remain largely unaffected by integrability breaking, while other properties are dramatically altered. For typical experimental parameters \cite{Nguyen_solcoll_controlled}, soliton shape \cite{Khaykovich_quinticsol_PhysRevA} and fragmentation time-scales \cite{streltsov_frag} are among the former, and different aspects of collision dynamics among the latter. A major qualitative change of dynamics in Q1D arises through new collision channels, in which atomic population can transfer between the colliding solitons. Extending \rref{aparna_collisions}, our results explicitly take into account integrability breaking interactions due to finite transverse extent of the Bose gas, and employ the truncated Wigner approximation (TWA) \cite{steel:wigner,Sinatra2001, castin:validity,blaire:review} to represent the post-collision state sketched in \fref{sketch}. 

As we shall show, in the emergent picture of a colliding pair of solitons these first loose mutual phase coherence in the ensemble, only to then strongly entangle (see also \rref{aparna_hyperent}) through atom number changing collisions. Disappearing visibility of quantum interference in averages and ubiquitous entanglement generation between a quantum system and its environment are both crucial for a classical appearance arising from quantum mechanics
\cite{Schlosshauer_decoherence_review}. 
Bright condensate solitons appear a versatile probe to explore this behavior of matter at the quantum-to-classical boundary as their ensemble average exhibits both, tuneable decoherence and tuneable generation of mesoscopic entanglement, while the density evolution
in each single realisation of the ensemble remains observable through in-situ imaging \cite{Nguyen_solcoll_controlled}.

\section{Soliton collisions}

Collisions of inert classical objects are typically fully governed by initial positions and momenta of collision partners. Quantum mechanically, collisions might additionally depend on the quantum phases in the many-body wave function. The latter also play a central role in condensate soliton collisions, which are controlled by the relative phase $\varphi$ and distance $d$ between the colliding solitons in mean field theory.  We write a twin soliton mean-field wave function as
\begin{align}
\phi_{0}(x) &=l(x)e^{ikx} +e^{i\varphi} r(x)e^{-ikx},
\label{soliton_pair_meanfield}
\end{align}
with left and right soliton shapes $l(x)={\cal N} \:\: \mbox{sech}[(x+d/2)/\xi]$, $r(x)={\cal N} \:\: \mbox{sech}[(x-d/2)/\xi]$, where ${\cal N}$ normalises each soliton to contain
$\sub{N}{sol}=\int \upd x |l(x)|^2=\int \upd x |r(x)|^2$ constituent atoms. The soliton widths are set by the healing length scale $\xi$, while solitons are a distance $d$ apart with relative phase $\varphi$. $k$ is the wave number associated with symmetric bulk soliton motion.

In 1D mean field theory, solitons evolve by the GPE, 
\begin{align}
\label{GPE}
i \hbar\frac{\partial}{\partial t}\phi(x,t) =  \left[ -\frac{\hbar^2}{2m} \frac{\partial^2}{\partial x^2}  + \tilde{g}_{1D}{|\phi(x,t)|}^2\right] \phi(x,t),
\end{align}
with 1D interaction strength $\tilde{g}_{1D}=2a_s(\hbar\omega_\perp)=\sub{U}{0}/(2\pi\sigma_{\perp}^2)<0$ derived from the 3D interaction strength $\sub{U}{0}=4 \pi \hbar^{2} a_{s}/m$ and a transverse oscillator length $\sigma_{\perp}=\sqrt{\hbar/(m\omega_\perp)}$. Here $\sub{a}{s}$ is the scattering length, $\omega_\perp$ the transverse trapping frequency and $m$ the atomic mass. The soliton width is then $\xi=2\hbar^2/(m |\tilde{g}_{1D}| \sub{N}{sol} )$.
For a simplified description, the Ansatz \bref{soliton_pair_meanfield} can be inserted into \bref{GPE} to derive effective equations of motion for $d(t)$ and $\varphi(t)$, predicting attractive collisions for $\varphi=0$ and repulsive collisions for $\varphi=\pi$  \cite{gordon_forces,stoof_solitons}.

\section{Beyond mean-field in the quasi-1D regime} 

Here we go beyond \eref{GPE}, taking into account the effective modification of interactions by transverse modes in an elongated Q1D trap, and incorporating quantum correlations beyond mean field theory.
It has been shown in \cite{Muryshev_darksolelong_quintic_PRL,Sinha_solfriction_quintic_PRL,Mazets_breakinteg_PhysRevLett}, that the former give rise to an effectively one dimensional Hamiltonian
\begin{align}
\hat{H}&= \int dx \bigg\{ \hat{\Psi}^\dagger
(x)\left[-\frac{\hbar^2}{2m}\frac{\partial^2}{\partial x^2} + \frac{\tilde{g}_{1D}}{2} \hat{\Psi}^\dagger(x)\hat{\Psi}(x)\right]\hat{\Psi}(x)
\CR
&
-\frac{\tilde{g}_2}{3} \hat{\Psi}^\dagger(x)\hat{\Psi}^\dagger (x) \hat{\Psi}^\dagger (x)\hat{\Psi}(x)\hat{\Psi}(x) \hat{\Psi}(x)\bigg\}.
\label{1DHamil_threebody}
\end{align}
The field operator $\hat{\Psi}(x)$ annihilates an atom of mass $m$ at the longitudinal position $x$. Effective three body interactions scale as $\tilde{g}_{2}=U_\perp/(3\pi^{2}\sigma_{\perp}^4) > 0$ with $U_\perp=72 \ln(4/3)\hbar^{3} a_{s}^{2} \pi^{2}/(m^{2} \omega_{\perp})$ \cite{Mazets_breakinteg_PhysRevLett}. 

Using the usual techniques \cite{steel:wigner}, we find the truncated Wigner equation of motion for the stochastic wavefunction $\phi_W(x,t)$ in \eref{TWA3B} in a dimensionless form, by rescaling the wavefunction as $\phi_W \rightarrow \phi_W \sqrt{D}$, space as $x\rightarrow x/D$ and time as  $t\rightarrow t/T$ where $D=\sigma_\perp$ and $T=\omega_{\perp}^{-1}$:
\begin{equation}
\begin{split}
 \label{TWA3B}
i \frac{\partial}{\partial t}\phi_W=  \bigg[-\frac{1}{2} \frac{\partial^2}{\partial x^2} + g_{1D}{(|\phi_W|}^2 - \delta_c)&\\ -q_{2}\big(|\phi_W|^4-2|\phi_W|^2  \delta_c+ \delta_c^{2}\big) \bigg]\phi_W.
\end{split}
\end{equation}   
The dimensionless interaction constants in \bref{TWA3B}  are $g_{1D}=2 a_{s}/\sigma_{\perp}$, and $q_{2}=24 \ln[\frac{4}{3} ]a_{s}^{2}/\sigma_\perp^2$,
while $\delta_c=\delta_c(x,x)\approx1/dx$ is a commutator \cite{norrie:long} for finite grid-spacing $dx$.

To first discuss field theory, we drop $\delta_c$ and replace $\phi_W(x)$ by the mean-field $\phi(x)$. Due to the quintic term, soliton shapes change from $l(x)\rightarrow L(x)$ and $r(x)\rightarrow R(x)$ in \eref{soliton_pair_meanfield} such that~\cite{Khaykovich_quinticsol_PhysRevA}
\begin{align}
\label{quintic_solmodes}
R(x,t) = \sqrt\frac{-4 \mu \sqrt{3/(4 q_2)}}{\sqrt {g^{2}-4 \mu }\cosh[2\sqrt{-2 \mu} (x-d(t))]+g},
\end{align}
using $g=-\sub{g}{1D}  \sqrt{3/(4\sub{q}{2})}$,
and $L(x,t)$ similarly with $d(t)\rightarrow -d(t)$. The number of atoms in a soliton $\sub{N}{sol}= \sqrt \frac{6}{\sub{q}{2}} \arctan\bigg[\frac{2 \sqrt {-\mu}}{g+\sqrt{g^{2}-4\mu}}\bigg]$
is controlled by the chemical potential $\mu<0$. Soliton shapes are compared in \frefp{fragmentation}{a} for two different quintic interaction strengths $\sub{q}{2}$.

In the truncated Wigner approximation, we augment the initial state \bref{soliton_pair_meanfield} with the above quintic soliton modes to a stochastic field $\phi_{W}(x)$,
 through the prescription 
 $\phi_{W}(x,0) = \phi_0(x) + \frac{1}{\sqrt 2} \zeta(x)$, 
where $\phi_0(x)$ is the initial mean field wavefunction and $\zeta(x)$ is a complex Gaussian distributed random function with correlations $\overline{\zeta(x)\zeta(x')}=0$ and $\overline{\zeta^*(x)\zeta(x')}=\delta_c(x,x')$. The overline denotes the stochastic average. 
Quantum correlations are found through stochastic averages such as \cite{book:qn,blaire:review}
\begin{equation}
\label{averages}
\expec{\hat{\Psi}^\dagger(x)\hat{\Psi}(x')}=\overline{\phi_{W}^*(x)\phi_{W}(x')}-\delta_c(x,x')/2,
\end{equation}
where the field operator $\hat{\Psi}(x)$ annihilates an atom at $x$. For atoms with cubic nonlinearity, the TWA is valid for short times and strong mean field \cite{blaire:review}, covering the crucial moment of collision here. See \cite{Drummond_waveguide,drummond_measure} for some early applications of the TWA to solitons and their collisions. We assume quintic nonlinearities do not strongly alter its validity.

For analytical insight, we will also consider a two-mode model (TMM) that arises from \eref{1DHamil_threebody} by insertion of the ansatz $\hat{\Psi}(x,t) =\overline{L}[x,d(t)] \hat{a}(t) + \overline{R}[x,d(t)]  \hat{b}(t)$ for the atomic quantum field, where $\hat{a}$ destroys a boson in the left soliton, with \emph{mode} function $\overline{L}(x,t) =L(x,t)/\sqrt{\sub{N}{sol}}$, and $\hat{b}$ does the same for the right soliton. Each atom can thus be either in the left or the right soliton. The mode functions depend on time through the inter-soliton separation $d(t)$. Inserting the Ansatz into \bref{1DHamil_threebody} and assuming large $d(t)$ so that modes $\overline{L}(x)$ and $\overline{R}(x)$ do not overlap, we reach a simple TMM Hamiltonian ($\hbar=m=1$)
\begin{align}
\label{TMM_Hamil}
\hat{H} &= \omega(\hat{a}^{\dagger}\hat{a}+ \hat{b}^{\dagger}\hat{b})+ \frac{\chi}{2} ( \hat{a}^\dagger \hat{a}^\dagger \hat{a} \hat{a} + \hat{b}^\dagger \hat{b}^\dagger \hat{b} \hat{b}) \CR 
&+\frac{\eta}{3}(\hat{a}^{\dagger}\hat{a}^{\dagger}\hat{a}^{\dagger}\hat{a}\hat{a}\hat{a}+\hat{b}^{\dagger}\hat{b}^{\dagger}\hat{b}^{\dagger}\hat{b}\hat{b}\hat{b}).
\end{align}
In \bref{TMM_Hamil}, $\omega=\int dx |(\partial/\partial x)  \bar{L}(x)|^2/2$ are single atom energies, 
and $\chi= \sub{g}{1D} \int dx |\bar{L}(x)|^4$, $\eta= -\sub{q}{2} \int dx |\bar{L}(x)|^6$ capture the strength of interactions. 

\section{Three-body contribution to phase diffusion and mode shape} 

Whether the two soliton modes are phase coherent can be inferred from the eigenvalues $(2\sub{N}{sol})\bar{\lambda}_{\pm}$ of the one-body density matrix (OBDM) \cite{penrose_onsager_crit}
\begin{align}
\label{modes_obdm}
\varrho= \begin{bmatrix}
\expec{\hat{a}^\dagger\hat{a}}  & \expec{\hat{b}^\dagger\hat{a}} \\
\expec{\hat{a}^\dagger\hat{b}}   & \expec{\hat{b}^\dagger\hat{b}} 
\end{bmatrix}.
\end{align}
If $\varrho$ has one dominant eigenvalue $\bar{\lambda}_{+}\approx 1$, all the atoms reside in the same \emph{single} particle state (orbital), which can represent \emph{two solitons} with complete phase coherence as in \bref{soliton_pair_meanfield}. Otherwise the system is fragmented with no phase-coherence between solitons \cite{streltsov_frag,mueller:fragmentation}, hence $\expec{\hat{a}^\dagger\hat{b}}=0$. 
It was earlier shown that soliton trains fragment \cite{streltsov_frag}, which we could trace back to phase diffusion \cite{lewenstein_phasediff,menotti:splitting,soliton_phasediff_haine} in \rref{aparna_collisions}. 
We now explore if this picture is modified through additional three-body interactions.

\begin{figure}[htbp]
\includegraphics[width=1.0\columnwidth]{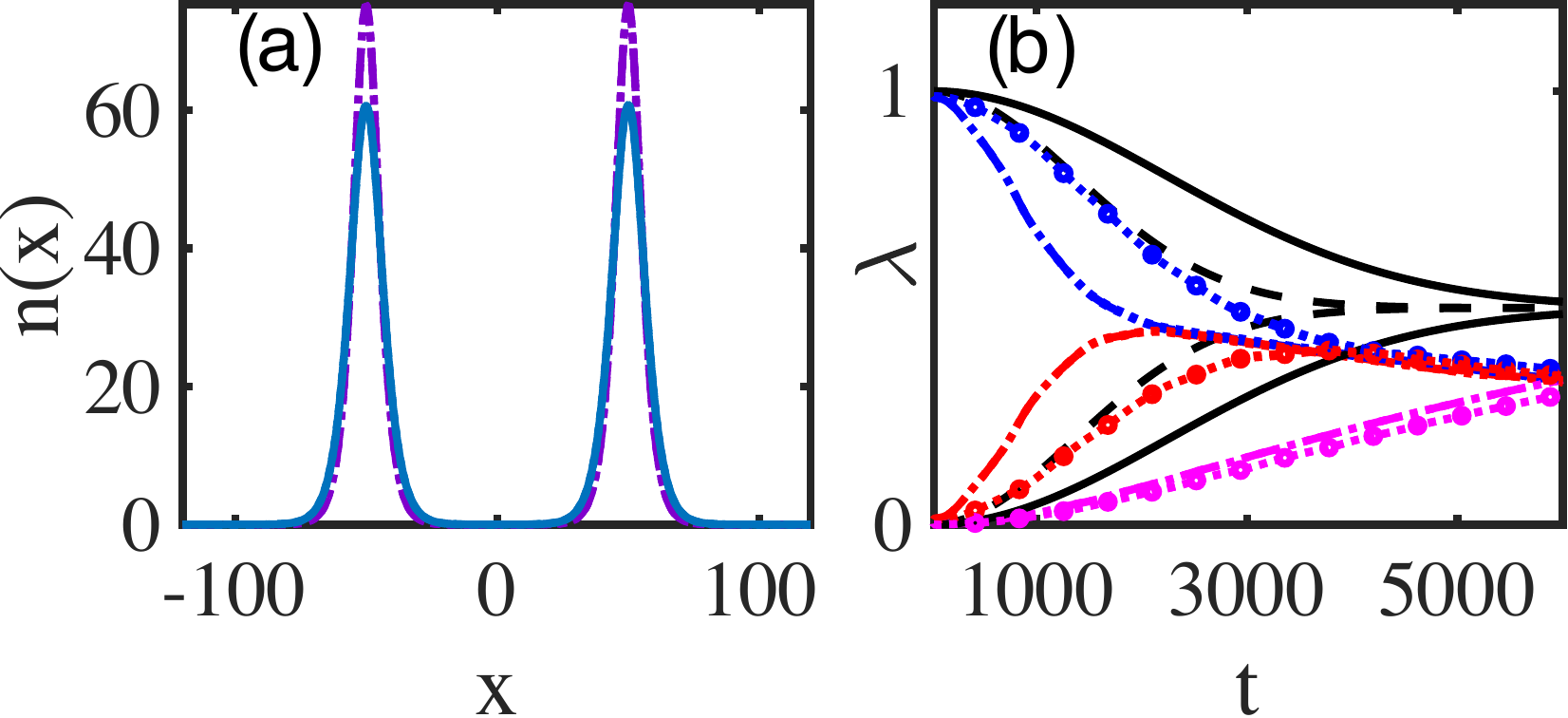}
\caption{\label{fragmentation} Soliton shape and fragmentation of bright BEC solitons with quintic non-linearity, for $\sub{N}{sol}=1000$, $g_{1D}=-2.3\times10^{-4}$ corresponding to a scattering length $a_s=-0.15$ nm and $\omega_\perp/(2\pi)=800$ Hz (a) The mean atomic density $n(x)=\expec{\hat{\Psi}^\dagger(x)\hat{\Psi}(x)}$ from TWA using \eref{averages}, of a soliton pair, for two different quintic nonlinearities (light blue solid $\sub{q}{2}=\sub{q}{2a}=9.6 \times 10^{-8}$, violet dot-dashed $\sub{q}{2}=\sub{q}{2b}=7.68\times 10^{-7}$). For $q_2=0$, the shape is indistinguishable from the light blue line. (b) Relative occupation $\bar{\lambda}_k$ of all system orbitals (eigenvalues of \bref{modes_obdm}) from TWA  (colored lines) and $\bar{\lambda}_{\pm}$ from TMM (black lines). We again compare $\sub{q}{2a}$ (marked with $\circ$ for TWA, black solid for TMM) and $\sub{q}{2b}$ (without marker for TWA, black dashed for TMM). Blue indicates the initially occupied orbital and red the second orbital participating in fragmentation. Magenta shows the sum of all other orbital populations in TWA. The sampling error for all TWA results is not visible. 
}
\end{figure}
For two far separated solitons, such that \bref{TMM_Hamil} is valid, we can evaluate the OBDM time-evolution, starting from both solitons in a coherent state with mean number $\sub{N}{sol}$ corresponding to a pure BEC. We choose this state for simplicity and its natural connection to mean field theory. In experiments, the initial state will depend strongly on the soliton preparation protocol \cite{billiam_bmw,Edmonds_brightsol_noisefree,Maxim_breather,Maxim_fluctuation}. The solutions allow us to extract the relative occupation $\bar{\lambda}_{\pm}$ of the two system orbitals and hence degree of fragmentation as
\begin{align}
\bar{\lambda}_{\pm}&=\left(1  \pm e^{2 \sub{N}{sol} \left[\cos(t/(2 (\sub{N}{sol}-1) \eta +\chi))-1 \right]}\right)/2,
\label{eigenvalues}
\end{align}           
which becomes $\bar{\lambda}_{\pm}\approx \left(1\pm e^{-[t/\sub{t}{frag}]^2}\right)/2$ at short times. The system thus fragments on the timescale 
\begin{align}
\sub{t}{frag}=\left|\sqrt{\sub{N}{sol}}(2 (\sub{N}{sol}-1)\eta + \chi)\right|^{-1}.    
\label{Frag_Time}
\end{align}
For negligible quintic interactions, \eref{Frag_Time} reduces to the fragmentation time of the cubic model \cite{aparna_collisions}. We see that in a general two-mode system, fragmentation can be accelerated or delayed, depending on the relative sign of cubic or quintic interactions. However for bright solitons $\chi,\eta<0$, hence here the fragmentation process must be accelerated. 

The TMM roughly agrees with the substantially more involved TWA regarding this time-scale and its dependence on $\eta$, as shown in \frefp{fragmentation}{b}. 
For this comparison, we extract the OBDM $\varrho(x,x')=\expec{\hat{\Psi}^\dagger(x)\hat{\Psi}(x')}$ from the TWA simulation and diagonalise it as a function of time, yielding eigenvalues $\lambda_k(t) =(2\sub{N}{sol})\bar\lambda_k(t)$. Initially, we have a pure BEC of two solitons since $\bar{\lambda}_+=1$. If two $\bar\lambda_{k}$, are of the order of unity, the system is fragmented. The figure shows that fragmentation occurs faster as the strength of the quintic non-linearity increases. The chosen examples show a considerable acceleration of fragmentation by stronger quintic interactions. However for parameters corresponding to recent experiments \cite{Nguyen_solcoll_controlled}, we would have $(\sub{N}{sol}-1)\eta =-8.86\times10^{-7}$ compared to $\chi=-3.89\times10^{-6}$, thus the quintic contribution in \eref{Frag_Time} would have only a minor impact there. Our assumption of an elongated Q1D trap and the validity of the quintic correction model \bref{TMM_Hamil} require $\sub{N}{sol}\eta < \chi$, thus the fragmentation through phase diffusion can only appear mildly accelerated in general. 

\section{Atom transfer in soliton collisions} 

In contrast to the minor quantitative impact of three-body interactions on fragmentation times, they can cause qualitatively new features in quantum soliton collisions.
To show this, we now study the interplay of fragmentation and collisions, using TWA.
\begin{figure}[htb]
\centering
\includegraphics[width=0.89\columnwidth]{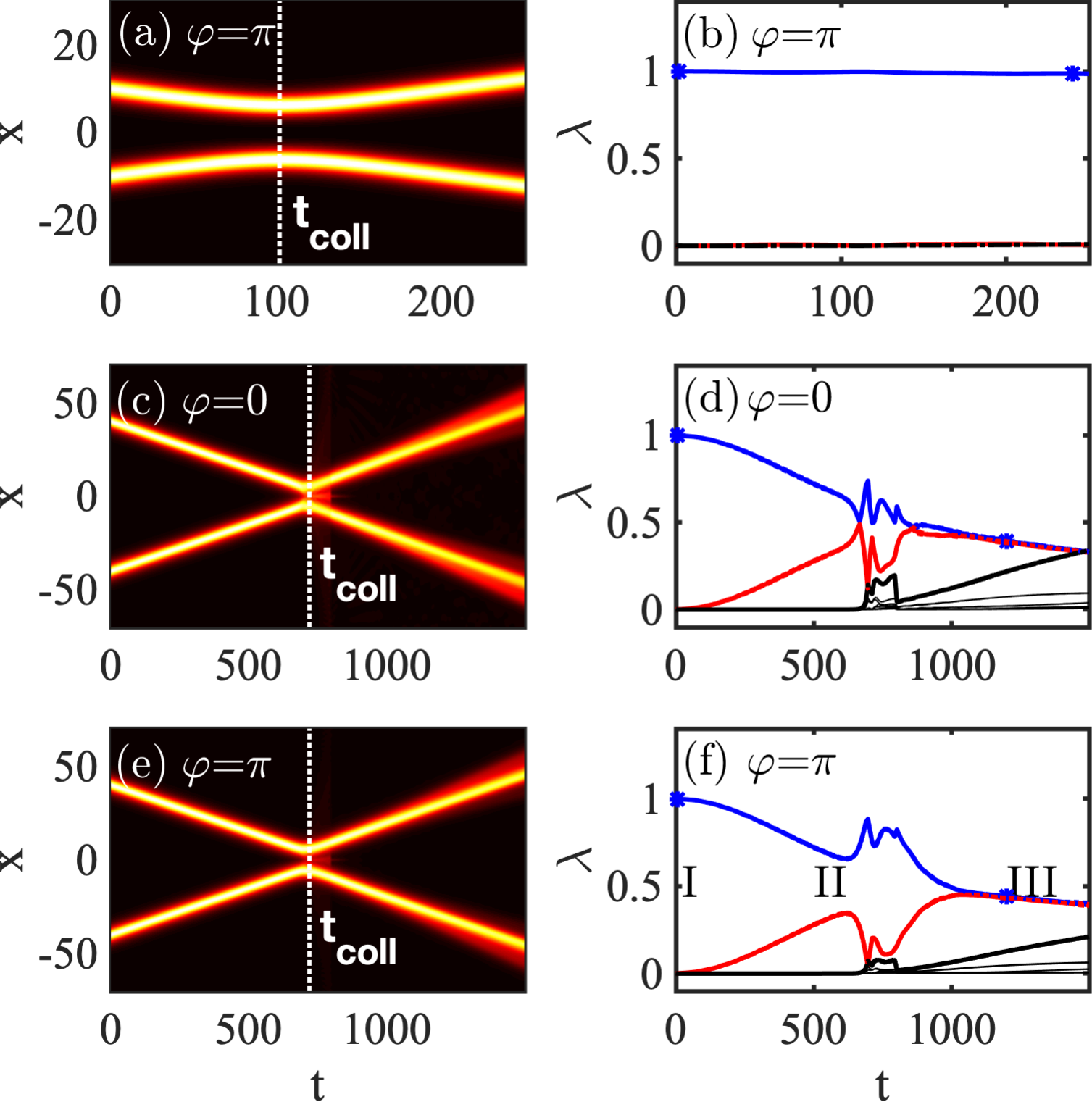}
\caption{\label{collisions} Collision and coherence dynamics in controlled soliton collisions with three-body interactions, before fragmentation, at time $\sub{t}{coll}<\sub{t}{frag}$ (a-b) and after fragmentation, at time $\sub{t}{coll}>\sub{t}{frag}$ (c-f), with the same initial velocity $\sub{v}{ini}\approx 0.05$. We use $\sub{N}{sol}=28000$, $g_{1D}=-2.53\times10^{-5}$ and $q_2=1.10\times10^{-9} \equiv \bar{q}_{2}$ unless otherwise indicated, corresponding to a scattering length $a_s=-0.030$ nm and $\omega_\perp/(2\pi)=254$ Hz, $D=2.38$ $\mu$m, $T=0.62$ $m$s. The initial relative phases between solitons, $\varphi$, are indicated. (a,c,e) Mean atomic density $n(x)=\expec{\hat{\Psi}^\dagger(x)\hat{\Psi}(x)}$ from TWA in (a) and square root of mean atomic density $\sqrt{n(x)}$ to emphasize weak features in (c) and (e). (black, zero; bright, high). (b,d,f) The two largest orbital populations $\bar{\lambda}_k(t)$ from TWA (solid blue and red). The remaining populations are shown as thin black lines (two states per line), and their sum as a thick one. (f) Roman numbers refer to the three regimes shown in \fref{sketch} and discussed in the conclusion. Sampling errors are not visible on this scale. }
\end{figure}
We separate the fragmentation and collision time-scales by forcing solitons to collide at a set time $\sub{t}{coll}=|d/(2\sub{v}{ini})|$, where $d$ and $\sub{v}{ini}$ are their initial distance and velocity.  Since the noise added in the initial state also causes an uncertainty of the soliton centre of mass (CM) and velocity \cite{Cosme_com_motion,weiss_CMdiffusion}, the collision time $\sub{t}{coll}$ becomes uncertain. 
This is a distraction from our focus on the collision itself, hence we remove CM diffusion by post-processing the noisy initial state as discussed in \rref{aparna_hyperent}. Subsequent to this step, solitons collide at the target time $\sub{t}{coll}$ for all realisations of the noise $\zeta(x)$. 
We also process time-evolving trajectories to remove those for which the number imbalance reaches $2a(t)=n_L(t)-n_R(t)>10000$ from averages such as \bref{averages}. This is to eliminate breathers and mergers as discussed in \rref{aparna_hyperent} and focus on collisions with a binary final state. In the expression above, $n_L(t)=\int_{-\infty}^0 dx\:  n(x,t)= \int_{-\infty}^0 dx\:[ \overline{|\phi_{W}(x)|^2}-\delta_c(x,x)/2]$ is the total atom number on the left side of the spatial domain, $n_R$ is the total number on the right.

The TWA now allows a more complete study of soliton motional and fragmentation dynamics than the methods employed \rref{aparna_collisions}, adding the effective three-body interactions $\sim q_2$. The colour shading in \fref{collisions} (a,c,e) indicates the mean atomic density $n(x)=\expec{\Psi^\dagger(x)\hat{\Psi}(x)}$, obtained as in \eref{averages}, averaging over $\sub{N}{traj}=20000$ trajectories. As discussed earlier, TWA also provides us the time evolution of the eigenvalues $\bar\lambda_k$ of the OBDM, shown in panels (b,d,f). 

\begin{figure}[htb]
\centering
\includegraphics[width=0.8\columnwidth]{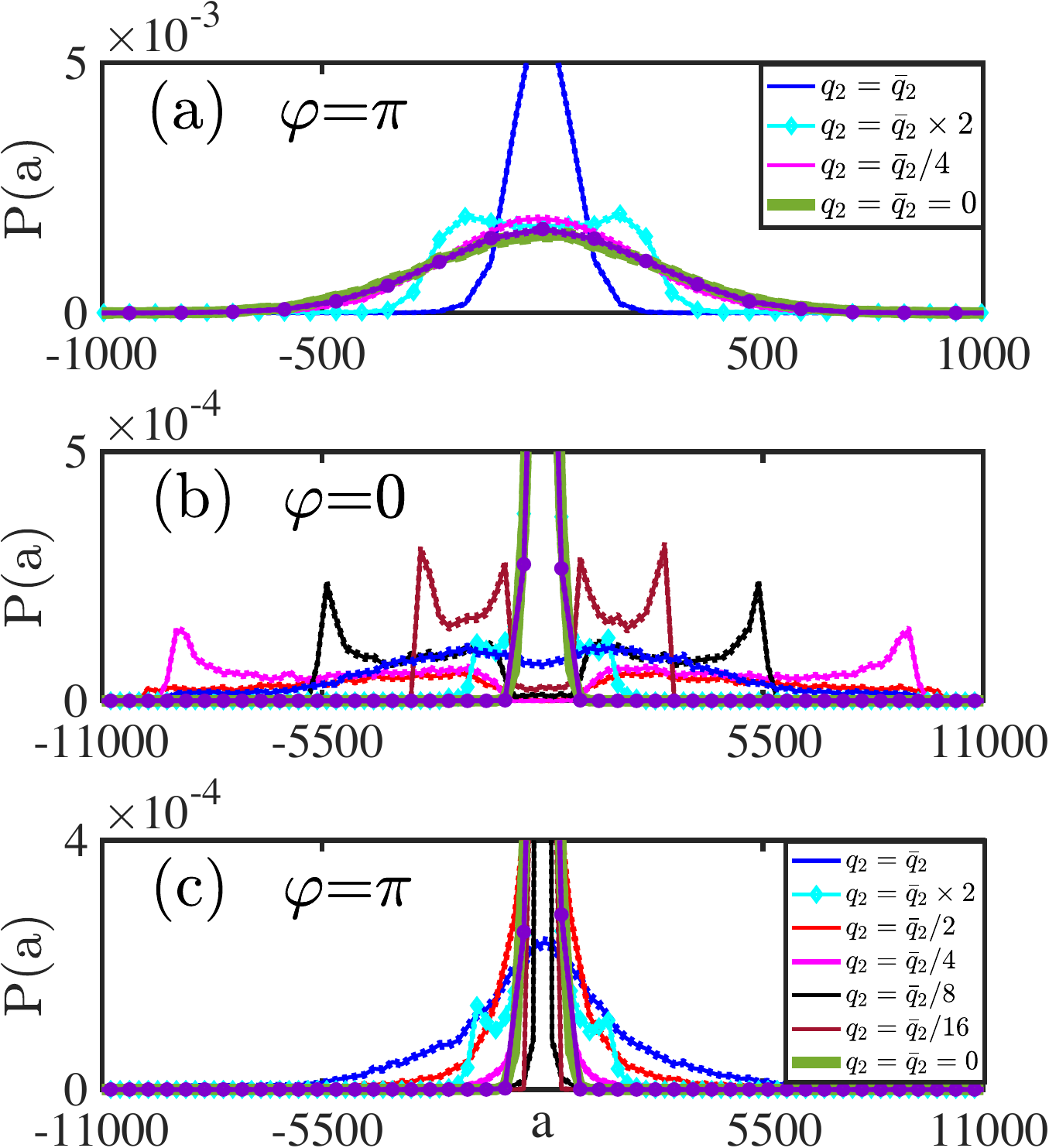}
\caption{The probability distributions for the atom number difference  $2a=n_L-n_R$ 
 from the TWA before collisions (violet lines with $\bullet$) and after collisions (other colours), for parameters as in \fref{collisions}, near experiments in \rref{Nguyen_solcoll_controlled}, which set $\bar{q}_2$. We then vary the strength of three-body interactions $q_2$ relative to the reference value $\bar{q}_2$.
Dotted lines close to solid lines indicate the sampling error. We compare collisions before fragmentation in (a) with those after fragmentation in (b,c). The thick green line in (a-c) is the post-collision distribution for vanishing quintic term, $\bar{q}_2=0$. The snapshots are at the times marked by (blue $\star$) in \fref{collisions} before the collision (here violet lines with $\bullet$) and after the collision (here other lines).
\label{Probability_Distribution}
 }
\end{figure}
Prior to fragmentation, we find that solitons with $\varphi=\pi$ collide repulsively, as predicted by mean-field theory \bref{GPE}. The case $\varphi=0$ colliding before fragmentation is not shown in \fref{collisions} and  \fref{Probability_Distribution}, since it leads to a merger \cite{Khaykovich_quinticsol_PhysRevA} for these parameters, which may entail collapse in three dimensions \cite{Parker_BSW_coll,parker_bsw}. After fragmentation, the mean collision appears repulsive for all $\varphi$. A new feature, unlocked by three-body interactions, is the widened position uncertainty of solitons that collide after fragmentation. This can be seen upon close inspection of the mean atomic density in \frefp{collisions}{c,e}, which appears more diffused after collisions. The cause are inelastic collisions, in which the outgoing soliton velocity is not equal to the incoming one.

To see this, we inspect the atom number distribution within solitons denoted by $P(a)$. Here, we  encounter a major qualitative difference between collisions with and without three-body interactions, that constitutes our main result. \fref{Probability_Distribution} shows the distribution of the atom number difference $n_{L}-n_{R}$ before and after the collisions. For the case $q_2=0$, the distribution stays conserved, see thick green lines in panel (a). This feature is qualitatively changed by three-body interactions. While the distribution remains narrow for the initial relative phase $\varphi=\pi$ and collisions prior to fragmentation, in panel (a), it is widened for all initial relative phases by collisions after fragmentation, in panels (b) and (c). We compare this effect in \fref{Probability_Distribution} for various values of $q_2$,  relative to the reference value $\bar{q}_2$, representing the experimental parameters of \rref{Nguyen_solcoll_controlled}, and find a non-monotonic dependence of the distribution width on $q_2$, explained shortly.

Collisions with $q_2\neq 0$ are dramatically different from those with $q_2=0$ because the quintic nonlinear term
breaks the integrability of the TWA equations \bref{TWA3B}. While we had already reported atom transfer similar to \fref{Probability_Distribution} in \rref{aparna_collisions}, methods there broke integrability in an uncontrolled manner even for $q_2=0$. 
Here, the TWA captures both, integrability for $q_2=0$ and its breakdown through the same scattering terms that break it in soliton collision experiments of interest. Neither fragmentation nor the widening of the number distribution is quantitatively changed, if we include combined 1,2 and 3-body loss of about $1.78 \%$ of atoms while fixing $N(\sub{t}{coll})$, with methods discussed in \cite{yash_losshawking}.
  
We now explain why atom transfer in \fref{Probability_Distribution} is only prominent after fragmentation and not before. To this end, we use a phase-space picture for the single stochastic trajectories arising from \bref{TWA3B}, discussed in the next section.

 \section{Fragmentation enhances atom transfer}  \label{sec:frag}
\begin{figure}[htbp]
\includegraphics[width=1\columnwidth]{Fig5.pdf}
\caption{(a,b) Phase-space portrait (black lines) for population imbalance $z$ and inter-soliton phase difference $\varphi$, governing \esref{Coupled_DE1}{Coupled_DE2} for parameters guided by \fref{collisions} and \fref{Probability_Distribution}. We took a fixed distance $d=11$, the closest approach in \fref{collisions}. Then $\chi=-2.9 \times 10^{-6}$, $\eta=-1.8 \times 10^{-11}$, $\bar{U} = -4.3 \times 10^{-8}$, $\bar{T} = -1.8 \times 10^{-7}$, $\bar{K} = 2.2 \times 10^{-14}$, $\tilde{K} = 7.5 \times 10^{-14}$, $\tilde{J} = 9.4 \times 10^{-13}$ and $K=-0.0049$. Superimposed, we show (a) Trajectories corresponding to collisions before fragmentation for initial relative phase close to $\varphi=0$ (red $\bullet$ at $t=0$, blue $\bullet$ at $t=25$), and for initial relative phase close to $\varphi=\pi$ (teal $\bullet$ at $t=0$, brown $\bullet$ at $t=25$).
 (b) Trajectories corresponding to collisions after complete fragmentation with random initial relative phase (brown $\bullet$ at $t=0$, violet $\bullet$ at $t=25$, red $\star$ at $t=25$ with $q_2\rightarrow$ 2$\bar{q}_{2}$, for $\bar{q}_{2}$ defined in \fref{collisions}). Black dotted vertical lines at z=0 are a guide to the eye.
\label{Phase_Plots}
}
\end{figure}
To understand the different atom transfer probabilities before and after fragmentation, we move to the two-mode mean-field model for a soliton pair, 
writing $\phi(x,t) =\sub{\psi}{L}(t)\bar{l}(x,d_0)+ \sub{\psi}{R}(t)\bar{r}(x,d_0)$ for a representative fixed $d_0$, with time-dependent amplitudes $\sub{\psi}{L,R}(t) = \sub{\sqrt{N}}{L,R}e^{i \sub{\theta}{L,R}(t)} $, where $\overline{l}(x,t) =l(x,t)/\sqrt{\sub{N}{sol}}$, $\sub{{N}}{L,R}$ and $\sub{\theta}{L,R}$ are the number of atoms and phases of left and right soliton respectively. For the analytical results presented here, we use $l(x)$ for simplicity. We now insert this restricted ansatz into the mean field version of \eref{TWA3B} and following \cite{Smerzi}, rewrite the result in terms of the fractional population imbalance
$z =(N_L - N_R)/(2\sub{N}{sol})$ and inter-soliton phase difference $\varphi = \sub{\theta}{R}- \sub{\theta}{L}$, 
to reach equations of motion
\begin{subequations}
\label{Coupled_eqns}
\begin{align}
 \Dot{z} &=[f_{1}(z)\cos{(\varphi)}+f_{2}(z,\cos(2\varphi))]\sin{(\varphi)},
 \label{Coupled_DE1}
  \\
\Dot{\varphi} &= g_{1}(z)\cos{(\varphi)}+g_{2}(z)\cos(2\varphi)+g_3(z)\cos{(3\varphi)},
\label{Coupled_DE2}
\end{align}
\end{subequations}
where the functions $f_{k}$ and $g_{k}$ are given in the supplementary Material. Since \eref{Coupled_eqns} can be derived from an effective Hamiltonian as in \rref{Smerzi}, solutions for different initial conditions create a phase space portrait for the two dynamical variables $z$ and $\varphi$, shown in \fref{Phase_Plots} as thin black lines.

To understand our TWA results in \fref{Probability_Distribution}, we view the stochastic realisations of \bref{TWA3B} as noisy trajectories in this phase-space. When projecting onto the two-mode problem \eref{Coupled_eqns}, the noise $\zeta(x)$ contained in $\phi_W(x,t=0)$ causes a randomisation of the initial population imbalance $z(0)$ and initial relative phase $\varphi(0)$. This can be modelled by a swarm of trajectories in the phase space of \fref{Phase_Plots}, for which the initial distributions of $z(0)$ and $\varphi(0)$ roughly correspond to those in \fref{Probability_Distribution}. We extract a Gaussian fit for these two, $p(z)\sim e^{-z^2/(2 \sigma_z^2)}$ and similarly for $\varphi$, and find $\sigma_z=494.96/\sub{N}{tot}$ and $\sigma_\varphi=0.045 \: \pi$ for the distributions prior to collision in \fref{Probability_Distribution}.

We then use a corresponding ensemble of initial conditions as shown in \fref{Phase_Plots}(a), normally distributed around $z(0)=0, \phi(0)=0$ (red swarm of points) and $z(0)=0, \phi(0)=\pi$ (teal swarm of points), to understand how phase differences affect atom transfer prior to fragmentation. At a later time $t=25$, roughly corresponding to the duration of the collision in \fref{collisions}, we find that the trajectories starting near $z(0)=0$ and $\phi(0)=0$ move only slightly away from the unstable fixed point (blue points), while those starting near $z(0)=0$ and $\phi(0)=\pi$ remain close to the stable fixed point (brown points), as expected from the underlying phase space structure. Neither depart significantly from the line $z=0$, explaining why the pre-fragmentation scenario in \frefp{Probability_Distribution}{a} shows negligible atom transfer. To understand atom transfer after fragmentation, we follow a similar approach. A fragmented state can be described as a weighted average of coherent states  \cite{mueller:fragmentation}. To represent a completely unknown relative phase, we thus start with a uniform random distribution of all relative phase differences $\phi\in[-\pi,\pi)$, retaining the same distribution of $z(0)$ used earlier (brown swarm of points in \fref{Phase_Plots} (b)). Here we find a large fraction of the trajectories moving significantly away from the line $z=0$ by $t=25$ (violet points), due to the structure of phase space. This qualitatively explains the significant widening of the number distribution after fragmentation in \frefp{Probability_Distribution}{b,c}. 
We also capture the feature of \fref{Probability_Distribution}, that increasing $q_2$ does not necessarily lead to a wider final distribution, see red star points. This is due to cancellations among the TMM coefficients.

For the results in \fref{Phase_Plots}, the evolution times $t$ roughly correspond to the duration $\tau$ of collisions in \fref{collisions} and also the TMM parameters match that scenario. It is thus encouraging that the variances of distributions for $z$ obtained almost quantitatively match those shown in \fref{Probability_Distribution}. However one must bear in mind that the real soliton collision trajectory gives rise to a time-dependent $d(t)$ and thus time-dependent phase space structure, which will complicate the intuitive picture above. Coefficients in \eref{Coupled_eqns} can also change sign depending on $d$, which causes a swap of stable and unstable fixed points.

\section{Exploring the quantum to classical transition} 
  
The significant widening of the atom number distribution during a collision of bright solitons in \fref{Probability_Distribution} implies the generation of entanglement due to the number conservation of the underlying interactions \cite{Ng_nonloc_higherorder_PhysRevLett,aparna_hyperent}.
Further, if one soliton gains atoms at the expense of the other, momentum conservation requires it to move more slowly afterwards. The total post-collision many-body state then has the structure
\begin{align}
\ket{\Psi}&=\sum_{n_L} c_{n_L} \ket{n_L,v(n_L);n_R,v(n_R)},
\label{postcollstate}
\end{align}           
shown in \fref{sketch}. The states $ \ket{n_L,v(n_L);n_R,v(n_R)}$ in \eref{postcollstate} represent $n_L$ atoms in the left soliton, which moves with velocity $v(n_L)$, similarly for the right one. $c_{n_L}$ are coefficients, underlying the distribution $P(a)=|c_{a}|^2$ shown in \fref{Probability_Distribution}. The schematic \bref{postcollstate} constitutes a hyper-entangled soliton state, the realisation of which we explicitly demonstrate in \rref{aparna_hyperent} assuming pure states, using joint momentum and position uncertainties of solitons. 

\section{Conclusions and outlook} 
%
We have studied collisions of BEC bright solitary waves, explicitly including effective three-body interactions that break integrability in the Q1D setting \cite{Muryshev_darksolelong_quintic_PRL,Sinha_solfriction_quintic_PRL,Mazets_breakinteg_PhysRevLett}, for experimentally realistic parameters guided by \rref{Nguyen_solcoll_controlled}, using the the truncated Wigner approximation. While fragmentation due to phase-diffusion and soliton mode shapes are typically not significantly affected by these interactions for realistic parameters, we show that the collision dynamics changes qualitatively compared to the description neglecting these contributions: We demonstrate a significant change of the atom number distribution within each soliton by the collision due to three-body interactions, in contrast to the case of  two-body interactions where this distribution is conserved. This corresponds to a conversion of relative phase fluctuation into relative number fluctuations. Atom transfer may thus enable a direct probe of three-body interactions and phase distributions. We explained why fragmentation is a prerequisite for this atom transfer when starting from a repulsive relative phase $\varphi=\pi$, based on the phase-space structure of the noisy mean field model. 

The Bose gas many-body state passes through three stages, indicated in \fref{sketch} and \fref{collisions}, in which the OBDM shows a single (I), two (II) or many (III) occupied states. Firstly coherence between the solitons is lost through phase-diffusion, from I to II. Then, from region II to III,  quantum many-body dynamics in the collision leads to a state \bref{postcollstate} entangling atom numbers, positions and velocities of both solitons.  All three stages can be observed in experiments \cite{Nguyen_solcoll_controlled} and time-scales controlled. Also any entanglement degradation subsequent to the collision through losses or thermal scattering can be probed. Hence 
colliding bright solitons appear promising to gain a deeper understanding of the quantum-classical transition, owing to their relatively simple underlying Hamiltonian \bref{1DHamil_threebody}, yet complex emergent collisions.

\acknowledgments
We gladly thank the Max-Planck society for funding under the MPG-IISER partner group program. SK thanks Abhijit Pendse for discussions and ASR acknowledges the Department of Science and Technology (DST), New Delhi, India, for the INSPIRE fellowship IF160381.

\newpage
\onecolumn

\section{Supplementary Material}\centering
\section{Euler-Lagrange equations}

The non-linear two mode dynamical equations in terms of fractional population imbalance and inter-soliton phase difference are given by 
\begin{subequations}
\label{Coupled_eqns_simp}
\begin{align}
 \Dot{z} &=[f_{1}(z)\cos{(\varphi)}+f_{2}(z,\cos(2\varphi))]\sin{(\varphi)},
 \label{Coupled_DE_simp1}
  \\
\Dot{\varphi} &= [g_{1}(z)\cos{(\varphi)}+g_{2}(z)\cos(2\varphi)+g_3(z)\cos{(3\varphi)}]
\label{Coupled_DE_simp2}
\end{align}
\end{subequations}
in the main text. In detail the coefficient functions are given by
\begin{subequations}
\label{Coupled_eqns}
\begin{align}
f_{1}(z)&=2\sub{N}{tot}(\bar{U} + 2\tilde{K}\sub{N}{tot})(1-z^2),
\label{f1}
\\
f_{2}(z,\cos(2\varphi))&=-\sqrt{1-z^2}\bigg[2K - 2\sub{N}{tot} \bar{T} - \sub{N}{tot}^2
 \Big[
 \tilde{J}(1+z^2) +\bar{K}[2+\cos(2\varphi) ] (1-z^2)
 \Big]
 \bigg]
 \bigg\},
 \label{f2}
 \\
g_{1}(z)&=\frac{z}{2\sqrt{1-z^2}}\bigg\{
2\sub{N}{tot}\sqrt{1-z^2}\left(\chi  -2 \bar{U}-\sub{N}{tot} (\eta +3\tilde{K})\right)
\CR
&+\Big[4K-4  \sub{N}{tot} \bar{T} + \sub{N}{tot}^2\left[ \tilde{J} (2 - 6 z^2) - 9 \bar{K}  (1 - z^2)\right]
\Big]\bigg\},
\label{g1}
\\
g_{2}(z)&=-\frac{z}{2\sqrt{1-z^2}}\bigg\{2\sub{N}{tot}\left(2\sub{N}{tot}\tilde{K} + \bar{U} \right)\sqrt{1-z^2}\bigg\},
\label{g2}
\\
g_{3}(z)&=-\frac{z}{2\sqrt{1-z^2}}\bigg\{\bar{K}\sub{N}{tot}^2(1-z^2) \bigg\}.
\label{g3}
\end{align}
\end{subequations}

where $\sub{N}{tot}=2 \sub{N}{sol}$ and only $\eta$ and $\chi$, defined in the main text, are independent of the distance $d$. In contrast, defining the integral $O(\alpha,\beta)= \int dx\: \bar{l}(x)^\alpha \bar{r}(x)^\beta$, we have $\bar{U}=\sub{g}{1D} O(2,2)$,  $\bar{T}=\sub{g}{1D}O(3,1)$, $\tilde{K}=\sub{q}{2}O(4,2)$,  $\bar{K}=\sub{q}{2}O(3,3)$, $\tilde{J}=\sub{q}{2}O(5,1)$, 
which all depend on $d$, as does $K=\int dx (\partial/\partial x)  \bar{l}(x)(\partial/\partial x)  \bar{r}(x)/2$. $\chi$ and $\eta$ have the same form as in the main text, 
but with the replacement $\bar{L}(x) \rightarrow \bar{l}(x)$ and $\bar{R}(x) \rightarrow \bar{r}(x)$ for the application of \eref{Coupled_eqns_simp}.

\section{Soliton atom transfer in the Lieb-Liniger model}

Since the completely one dimensional soliton collision with $q_2=0$ can be studied using the analytical solution of the LL model \cite{McGuire_exactlysolvable_JMP,LL_model_PR} as in Refs.~\cite{lai_quantsol_I,lai_quantsol_II},
it is tempting to corroborate the unlocking of inter-soliton atom transfer by $q_2\neq 0$ also in that framework. 
We found that approach quickly intractable when trying to handle localized solitons, but could show the non-vanishing of the matrix element of three-body interactions between a state with two delocalized solitons with atom numbers $(\sub{N}{sol},\sub{N}{sol})$ and momenta $(p_0,-p_0)$ \cite{lai_quantsol_I,lai_quantsol_II} and one with numbers $(\sub{N}{sol}-k,\sub{N}{sol}+k)$ and momenta $(p_1',p_2')$ such that the total momentum is conserved.
\twocolumn

\bibliography{bibliography}

\begin{thebibliography}{10}
\bibitem{footnote_solitons}
The BEC solitons in question are solitary waves throughout this article.
\end{thebibliography}
\bibliographystyle{eplbib}
\end{document}